\documentclass[12pt]{article} 
\usepackage[koi8-r]{inputenc}
\usepackage[english]{babel}
\usepackage{pazh}
\usepackage{graphicx,amssymb,amsmath,natbib}
\usepackage{xspace}
\usepackage{grfext}

\usepackage{longtable}
\tightenlines
\usepackage{epstopdf}
\usepackage{authblk}

\newcommand{\Teff}{$T_{\rm eff}$}
\newcommand{\lgg}{log~$g$}

\newcommand{\eps}[1]{\log\varepsilon_{\rm #1}}
\newcommand{\kms}{$\rm km\ s^{-1}$}
\newcommand{\Eexc}{$E_{\rm exc}$}

\newcommand{\eu}[5]{\mbox{$#1\,^#2{\rm #3}^{#4}_{\rm #5}$}}
\newcommand{\Neltot}{$\rm [N_{el}/N_{tot}]$}

\newcommand{\Per}{$P$}
\newcommand{\vsini}{$V{\rm sin}i$}

\begin{document}




\title{HD 188101: A Spotted B Star with Abundance Anomalies}

\begin{center}
		R. M. Bayazitov$^{1*}$, L. I. Mashonkina$^{1, 2}$, Yu. V. Pakhomov$^1$, I. A. Yakunin$^3$ \\
		{\it
	    $^1$Institute of Astronomy, Russian Academy of Sciences, Moscow, 109017 Russia \\
		$^2$Institute of Laser Physics, Russian Academy of Sciences, Siberian Branch, Novosibirsk, 630090 Russia \\
		$^3$Special Astrophysical Observatory, Russian Academy of Sciences, Nizhnii Arkhyz, Karachai-Cherkessian
		Republic, 369167 Russia}
\end{center}

\vspace{2mm}
\vspace{2mm}

{\bf Abstract.} Based on spectroscopic and photometric observations, we have determined the fundamental
parameters of the poorly studied star HD 188101 with a weak magnetic field. Its effective temperature
$T_{\rm eff} = 14200 \pm 990$~K and surface gravity \lgg\ = 3.70 $\pm$ 0.16 are typical for main-sequence B9 stars. The
He, C, O,Mg, Si, Ti, and Sr abundances have been determined by taking into account the departures from
local thermodynamic equilibrium. Overabundances of Si, Ti, and Sr relative to their solar abundances have
been revealed. The He abundance is lower than the solar one, but the difference is within the error limits.
In addition to the photometric variability known from Kepler data, we have found changes in absorption
for He I, Mg II, Si II, Si III, Ti II, and Fe II lines, with different He I and Mg II lines giving different
abundances for the same phase of observations. The star HD 188101 is shown to belong to the group of
chemically peculiar He-weak SiTiSr stars.


\vfill
\noindent\rule{8cm}{1pt}\\
{$^{*}$ E-mail: rbayazitov@inasan.ru}


\section{INTRODUCTION}

About 10--30\% of the main-sequence (MS) B
and A stars are chemically peculiar (CP) \citep[][]{2007AstBu..62...62R, 2015ads..book.....M}. They show
overabundances or underabundances of chemical
elements relative to their solar abundances. According
to their kinematic characteristics, the CP stars
belong to the Galactic thin disk. They rotate around
their axis more slowly than do the normal stars.
The CP stars with a large-scale magnetic field and
periodic variations in brightness with an amplitude
of a few hundredths of a magnitude and spectral line
profiles belong to the class of Ap--Bp stars.

He-weak (or He-w) stars, in which the He I
lines are weaker than those in normal stars of the
same spectral type, are distinguished among the
CP stars \citep[][]{1971ApJS...23..213N}. About 30 He-weak stars
are known \citep[][]{2019MNRAS.487.5922G}.They are divided
into magnetic ones with an overabundance of silicon
(Si He-w) and/or titanium and strontium (TiSr Hew)
and nonmagnetic ones with an overabundance
of phosphorous and gallium (PGa He-w). Weak
magnetic fields \citep[][]{1983ApJS...53..151B} and, in some cases,
spectroscopic variability were revealed in the SiTiSr subgroup. The subgroup extends the sequence of
Ap stars to the range of high temperatures.

The theory of selective diffusion of chemical elements \citep[][]{Michaud1970} was proposed to explain the
observed abundance anomalies. In general, calculations
predict the gravitational settling of light elements
and the radiative expulsion of heavy elements.
The theory contains a large number of parameters
and, therefore, requires an observational check.

The star HD 188101 had not aroused any interest
of researchers until its light curves with an amplitude
of 0.02$^m$ obtained by the Kepler space observatory
were published \citep{2018A&A...619A..98H}. Such a
small amplitude of its brightness variations can be
explained by the presence of a spot (or spots) on
the surface. The star was included in the program
of a search for magnetic fields based on observations
with the Main Stellar Spectrograph (MSS) of
the Large Azimuthal Telescope (BTA), and a weak
magnetic field with a longitudinal component $B_z$ of
less than 1 kG was found in HD 188101 by \citet{2023AstBu..78..141Y}. Using the same spectra, \citet{2023AstBu..78..141Y} determined its effective temperature
$T_{\rm eff}$ = 14700~K, surface gravity log~g = 3.8, and rotational
velocity \vsini\ = 33 \kms\ and pointed out
that the Si II/Si III ionization balance is upset and it is impossible to describe the He~I 4471~\AA\ line profile
within the framework of a classical homogeneous
model atmosphere under the assumption of local
thermodynamic equilibrium (LTE).

Our preliminary analysis of the spectra for
HD 188101 showed that the star probably belongs
to the small group of He-weak stars that are distinguished
by a great variety of abundance anomalies. A detailed abundance analysis, with the abundance
determination for a large set of elements from various
lines at various ionization stages, is needed to
understand the chemical peculiarity mechanisms of
He-weak stars. Such information is available in the
literature only for very few objects, as can be seen from
the catalog data \citep{2019MNRAS.487.5922G}. Therefore,
in this paper we performed a detailed analysis of all
the available observational data for a probable new
member of the He-weak group, the star HD 188101,
to refine its atmospheric parameters, to determine
the abundances for the maximum possible number
of chemical elements, including, given the departures
from LTE, those for He, C, O, Mg, Si, Ti, and Sr, to
investigate the variability in lines and continuum, and
to establish the type of chemical peculiarity.

In Section~\ref{sect:obs} we present the sources of observational
data, the variability of the equivalent widths of
spectral lines, and the analysis of longitudinal magnetic
field measurements. The spectrum modeling
methods, including the construction of a model He I
atom, are described in Section~\ref{sect:methods}. The atmospheric
parameters and chemical composition are determined
in Sections~\ref{sect:atm} and \ref{subsecchabun}. Our results are discussed in
Section~\ref{sect:discus}. 

\section{OBSERVATIONAL DATA}\label{sect:obs}

\subsection{Distance, Photometry, and Spectra}

The latest Gaia data release \citep[Gaia DR3,][]{2021A&A...649A...1G} gives a parallax $\pi = 2.34\pm0.93$ (RUWE = 2.77) for HD 188101. RUWE (renormalized
unit weight error) is the factor that characterizes
the reliability of the Gaia parallax. For reliable values
RUWE should be less than 1.4. Despite the high
value of RUWE, we used the Gaia DR3 parallax.
Note that the parallaxes in the first Hipparcos catalogue \citep[$\pi = 2.31\pm0.77$,][]{1997ESASP1200.....E} and Gaia DR3 coincide within the error limits.

We used the photometric data\footnote{https://vizier.cds.unistra.fr/viz-bin/VizieR} \footnote{https://astroquery.readthedocs.io/en/latest/} in the range from 1000~\AA\ to 20 \micron\, obtained with various instruments\footnote{https://vizier.cds.unistra.fr/vizier/welcome/vizierbrowse.gml?bigcat} (Table~\ref{tab:filters}). To investigate the variability, we extracted the Kepler photometric data from the archive and
processed them following the instructions\footnote{https://lightkurve.github.io/lightkurve/tutorials/1-getting-started/using-light-curve-file-products.html?highlight=tesslightcurve}.

\begin{table*}[th]
		\centering
        \caption {The filters with their effective wavelengths and references to the catalogs} \label{tab:filters}
        \begin{tabular}{llll}\hline
                $\lambda_{\rm eff}$ (\AA) &  Filter             &        Catalog  &             Reference                      \\ \hline
                1565.54   &  TD1/WIDE.1565      & II/59B/catalog  &          \cite{1978csuf.book.....T}   \\
                1959.68   &  TD1/WIDE.1965      &                 &                                       \\
                2358.45   &  TD1/WIDE.2365      &                 &                                       \\
                2737.17   &  TD1/WIDE.2740      &                 &                                       \\
                4901.70   &  Hipparcos.Hp       & I/311/hip2      &       {\small \cite{1997ESASP1200.....E}}    \\
                5035.75   &  GAIA3.Gbp          & I/355/gaiadr3   &         \cite{2022yCat.1355....0G}    \\
                5822.39   &  GAIA3.G            &                 &                                       \\
                7619.96   &  GAIA3.Grp          &                 &                                       \\
                4810.16   &  PAN-STARRS/PS1.g   & II/349/ps1      &       \cite{2016arXiv161205560C}      \\
                6155.47   &  PAN-STARRS/PS1.r   &                 &                                       \\
                7503.03   &  PAN-STARRS/PS1.i   &                 &                                       \\
                8668.36   &  PAN-STARRS/PS1.z   &                 &                                       \\
                9613.60   &  PAN-STARRS/PS1.y   &                 &                                       \\
                12350.00  &  2MASS.J            & II/246/out      &      \cite{2003yCat.2246....0C}       \\
                16620.00  &  2MASS.H            &                 &                                       \\
                21590.00  &  2MASS.Ks           &                 &                                       \\
                33526.00  &  WISE.W1            & II/311/wise     &      \cite{2012wise.rept....1C}       \\
                46028.00  &  WISE.W2            &                 &                                       \\
                115608.00 &  WISE.W3            &                 &                                       \\
                220883.00 &  WISE.W4            &                 &                                       \\ \hline
        \end{tabular}
\end{table*}

We used the MSS spectra with a resolution $R = \lambda/
\Delta\lambda = 15000$ and a signal-to-noise ratio S/N = 150--200 in the spectral ranges 4420--4990~\AA\ or 4490--5055~\AA. We also used the spectrum taken with the
echelle spectrograph at the Nasmyth focus of BTA
(NES) with R = 40000 and S/N = 50 in the range 3950--6880~\AA. Table~\ref{tab:specs} gives a log of spectroscopic
observations. Here and below, the phases were
obtained using the ephemeris JD = 2455681.4953 + 3.98726E from \citet{2023AstBu..78..141Y}.

\begin{table*}[htbp]
	\centering
	\caption {Log of spectroscopic observations of HD 188101
		with MSS and NES} \label{tab:specs}
	\medskip
	\begin{tabular}{c|c|c|c}
		\hline
	Date & JD 245...& Rotation
	phase &  Range, \AA \\ \hline
		\multicolumn{4}{l}{MSS} \\
		03.04.19 & 8577.5409& 0.325 & 4436-4993\\
		26.04.19 & 8600.5416& 0.093 & 4494-5050\\
		16.05.19 & 8620.3812& 0.069 & 4498-5054\\
		17.05.19 & 8621.4375& 0.334 & 4498-5054\\
		20.05.19 & 8624.4812& 0.097 & 4498-5054\\
		20.10.19 & 8777.3298& 0.432 & 4425-4982\\
		21.10.19 & 8778.3361& 0.684 & 4425-4982\\
		11.11.19 & 8799.2812& 0.937 & 4425-4982\\
		14.11.19 & 8802.2541& 0.683 & 4425-4982\\
		17.11.19 & 8805.1486& 0.408 & 4425-4982\\
		05.06.20 & 9006.4236& 0.888 & 4425-4982\\
		29.07.20 & 9060.3083& 0.402 & 4425-4982\\
		30.07.20 &          &                  & \\
		04.09.20 &          &                  & \\
		\multicolumn{4}{l}{NES} \\
		12.10.19 & 8769.2721& 0.411& 3948-6981\\ \hline
	\end{tabular}
\end{table*}

\subsection{Line Intensity Variability}

In this section we investigate the variability of the
longitudinal magnetic field $B_z$ measured by \citet{2023AstBu..78..141Y} from the MSS spectra and the equivalent
widths of spectral lines.

To analyze the change in the longitudinal magnetic
field $B_z$ measured by \citet{2023AstBu..78..141Y}, we
used the model of a central magnetic dipole. It is
characterized by the magnetic field at the pole of the
dipole $B_p$ and the inclination of the dipole axis to the
stellar rotation axis $\beta$ (it changes from $0^{\circ}$ to $180^{\circ}$).
The inclination of the rotation axis to the line of sight
$i=42^{\circ}$ was calculated using the period of the photometric
variability P = 3.98726 days and \vsini\ from \citet{2023AstBu..78..141Y}. Positive values of $B_z$ are observed over
the entire rotation period and, therefore, the magnetic
pole cannot disappear from the visible stellar disk.
Taking $\beta=30^{\circ}$, we obtained an upper limit, $B_p = 3$ kG, allowing the atmosphere to be modeled without
taking into account the magnetic field (such magnetic
fields change the temperature in the atmosphere by
less than 100 K; \citep[][]{2005A&A...433..671K}). Note that
the observations are not described by the magnetic
dipole model; we estimate only the upper limit for $B_p$.

\begin{figure*}[th!]
	\centering
	\includegraphics[scale=0.7]{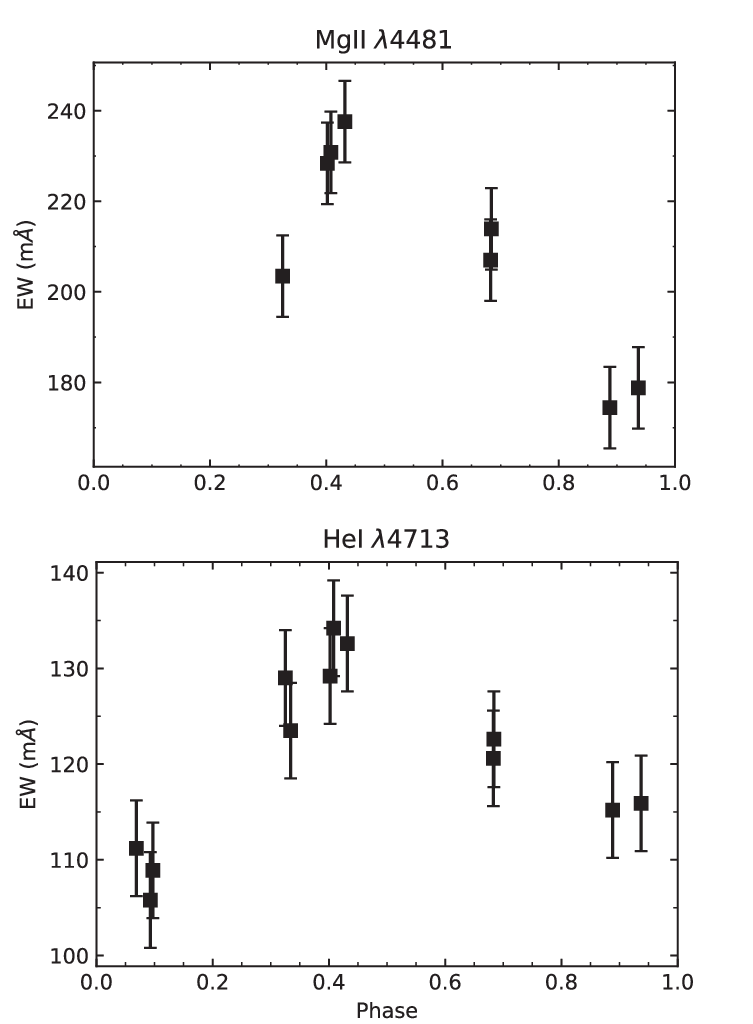}
	\caption{Changes in the equivalent widths of the Mg II and
		He I lines in HD 188101 with rotation phase.} \label{pic:varewmghe188}
\end{figure*} 

\begin{figure*}[th!]
	\centering
	\includegraphics[scale=0.7]{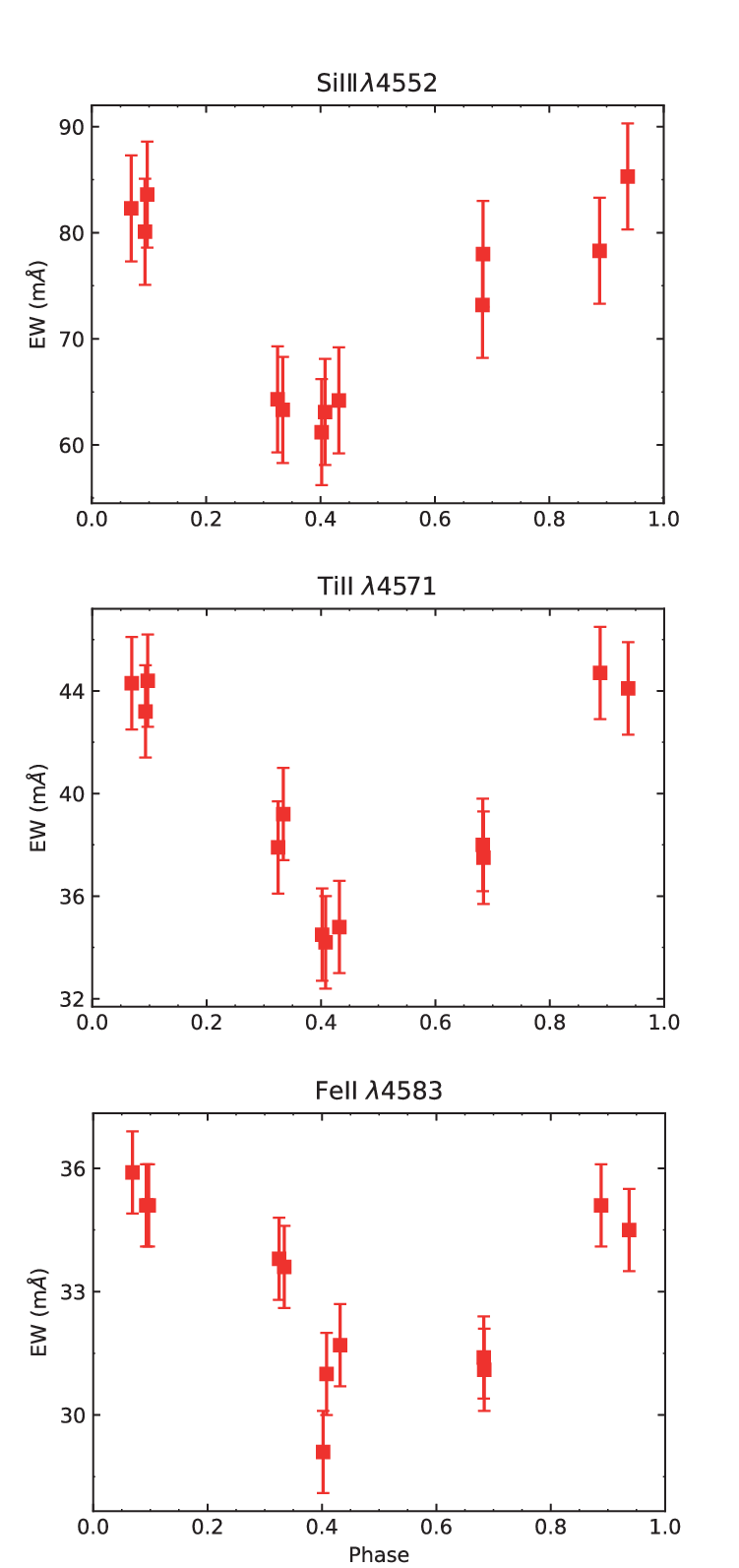}
	\caption{Changes in the equivalent widths of the Si III,
		Ti II, and Fe II lines in HD 188101 with rotation phase.} \label{pic:varewsitife188}
\end{figure*} 

Consider the variability in spectral lines. When the
MSS spectra are analyzed, the equivalent widths presented
in Figs. ~\ref{pic:varewmghe188} and \ref{pic:varewsitife188} give reliable information about
the variability. The strongest, unblended Mg II, He I,
Si III, Ti II, and Fe II lines are shown. The remaining
lines of these species exhibit a similar behavior. The
minima and maxima are seen to occur at phases 0.4
and 0.9. The changes in the Si II, Si III, Ti II, and
Fe II lines are out of phase with those in the He~I 4713~\AA\ and Mg~II 4481~\AA, lines, but are in phase with
the light curve.
We know peculiar stars in which an
anticorrelation of the absorptions in He I and Si II
lines \citep{1952ApJ...116..536D} and a correlation of the Mg and He abundances \citep[]{1993BSAO...36...52T}.
are
observed. The photometric variability and the change
in line absorption with stellar rotation phase point to
abundance inhomogeneities in the surface layers.

Note that we analyze the object as a single one.
First, within the limits of errors $\simeq$8 \kms, we found
no changes in the radial velocity. Second, the period
of the photometric variability, 4 days, is too short for
the star to be a binary.

\section{SPECTRUM MODELING METHODS}\label{sect:methods}

\subsection{Codes and Methods of Calculations}

The statistical equilibrium and radiative transfer
equations are solved using the DETAIL code \citep{Giddings81,Butler84,Przybilla2011}. The
model atom and the model atmosphere serve as the
input parameters; the level populations relative to the
LTE values serve as the output ones. Plane-parallel,
homogeneous model atmospheres were computed by
the LLmodels code \citep[line-by-line,][]{llmodels} with a detailed allowance for the absorption in lines
at a specified individual chemical composition of the
star. Table~\ref{tab:modat} lists the species whose lines we analyze
when abandoning LTE.

\begin{table}[h]
	\centering
	\caption{Model atoms} \label{tab:modat}
	\medskip
	\begin{tabular}{ll}
		\hline
		Atom    & Reference                \\ \hline
		He~I     & This paper              \\
		C~I-II   & \cite{2016MNRAS.462.1123A}\\
		O~I      & \cite{2013AstL...39..126S}\\
		Mg~I-II  & \cite{2018ApJ...866..153A}\\
		Si~I-III & \cite{2020MNRAS.493.6095M}\\
		Ti~I-II  & \cite{2020AstL...46..120S}\\
		Sr~II    & \cite{2020MNRAS.499.3706M}\\ \hline
	\end{tabular}
\end{table}

We determined the abundances of chemical elements
by the synthetic spectrum method, i.e., by
fitting the observed lines by theoretical profiles. The
synthetic spectrum was computed using the SynthVb
code \citep{synth}. For the lines of
the atom under study SynthVb uses the $b$-factors
computed by DETAIL; for the remaining lines we
assume LTE. For a specific level $b_i = n^{\rm NLTE}_i / n^{\rm LTE}_i$ characterizes the deviation of the population from the equilibrium one. Here, $n^{\rm NLTE}_i$ and $n^{\rm LTE}_i$ are
the level populations obtained when solving the
statistical equilibrium equations and from the Saha?Boltzmann formulas, respectively. The atomic data
needed to compute the synthetic spectrumwere taken
from the Vienna Atomic Line Database\footnote{https://vald.inasan.ru/~vald3/php/vald.php} \citep[VALD,][]{vald2015}.
The BinMag6 code \citep{Kochukhov_binmag} is used for the comparison with observations.

\subsection{Model He I atom}\label{sect:NLTE}

B--type stars are young objects, and the He abundance
in them should coincide with the cosmic one.
The protosolar value, $\eps{He}$ = 10.994$\pm$0.02 \citep{lodders21}, and the mean abundance for a sample
of B stars, $\eps{He}$ = 10.99$\pm$0.01 \citep[][]{Nieva2012}, may be considered as a standard cosmic
He abundance. Here, we use the scale $\eps{el} = {\rm lg}N_{\rm el}/N_{\rm H}$ + 12, where $N_{\rm el}$ and $N_{\rm H}$ are the number
densities of the atoms under study and hydrogen,
respectively. The deviation of the He abundance in a
star from the cosmic one is an indicator of the action
of chemical separation mechanisms in the surface
layers. To increase the accuracy of the He abundance
in the atmosphere of HD 188101, we used an approach
based on the abandonment of the assumption
of LTE (a non-LTE approach). For this purpose, we
constructed a model He~I atom.



\begin{figure*}[ht]
	\centering
	\includegraphics[scale=0.7]{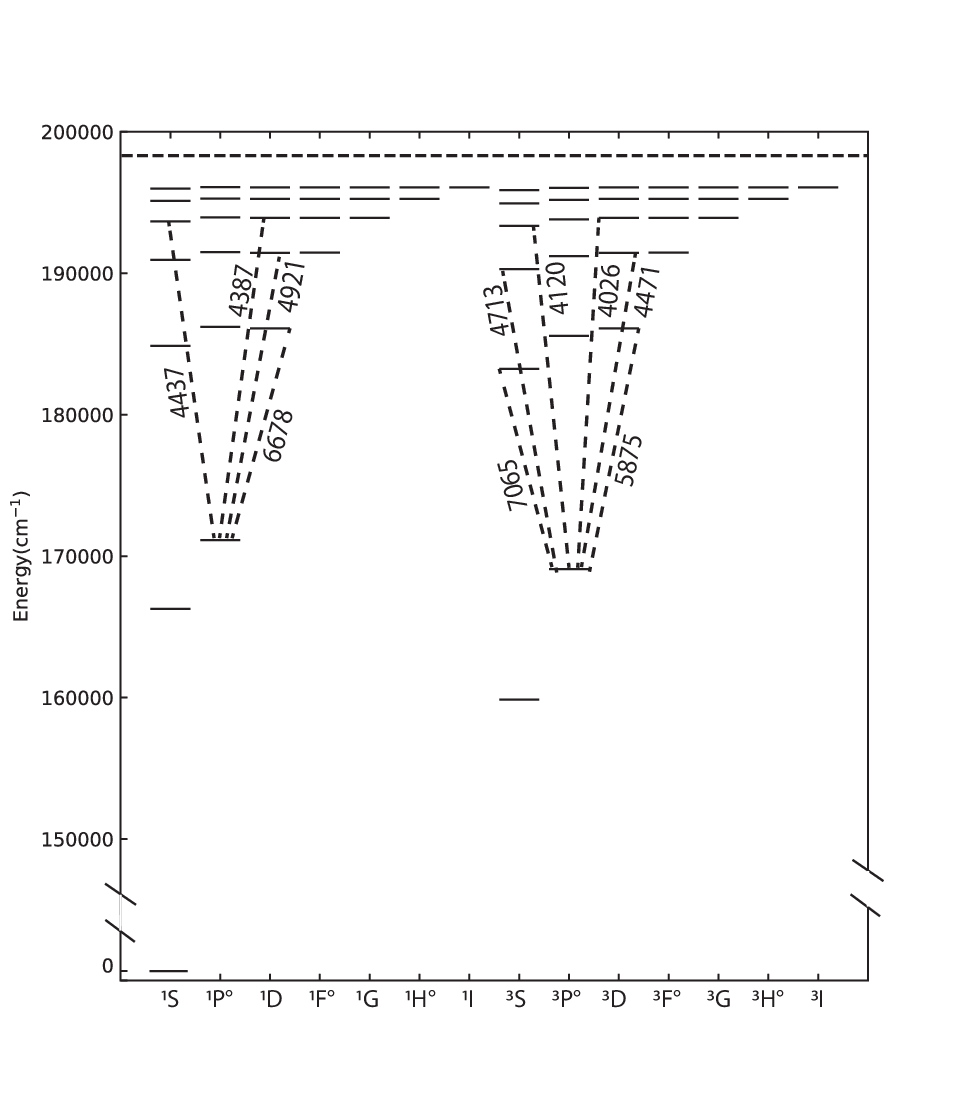}
	\caption{Energy levels in themodel He I atom. The dashes indicate the transitions in which the lines involved in determining the
		He abundance are formed (the wavelengths are in \AA).
	}\label{pic:LevelsHeI}
\end{figure*}

{\it Energy levels.} The model atom was constructed
using the singlet and triplet levels of the He~I 1snl electronic configurations with n $\le 7$ and l $\le 6$ from the NIST experimental database  NIST\footnote{https://www.nist.gov/pml/atomic-spectra-database} \citep{nist}. The system of He I levels was supplemented
by the ground He II state. The fine
splitting of the triplet states is ignored due to the small
energy separation, less than $10^{-4}$~eV. The uppermost
He I levels are separated from the continuum by 0.28~eV, which is less than the mean thermal energy
of electrons in the atmospheres of B stars, and
the electron-impact ionization and recombination
processes provide a close relationship between the
two helium ionization stages. The levels of the model
atom are shown in Fig.~\ref{pic:LevelsHeI}.


{\it Radiative transitions.} For the levels with an
orbital angular momentum L = S, P and D the
photoionization cross sections $\sigma_{ph}$(OP) calculated
within the Opacity Project were taken from TOPbase\footnote{https://cds.unistra.fr//topbase/topbase.html} \citep{topbase}. The \eu{np}{3}{P}{\circ}{} levels with n = 4, 5, 6 and 7, for which there are no data in TOPbase, are exceptions.
For the remaining levels we used the hydrogen-like photoionization cross sections.
For the levels with $\sigma_{ph}$(OP) our comparisons
showed that the hydrogen-like cross sections could
be close to $\sigma_{ph}$(OP), for example, for the \eu{2p}{3}{P}{\circ}{} level,
be larger than $\sigma_{ph}$(OP), for the \eu{2p}{1}{P}{\circ}{} level, and be
considerably smaller than $\sigma_{ph}$(OP), for the \eu{2s}{3}{S}{}{} level.

The photoionization cross sections for the He I
levels, including the np and nf levels absent in TOPbase,
were calculated by \citet{norad_he1} using the same
method as that in the Opacity Project, and the data
are presented inNORAD\footnote{https://norad.astronomy.osu.edu/}. Our comparisons showed
that the TOPbase and NORAD data agree for the
common levels. However, since $\sigma_{ph}$(NORAD) are
available for a larger set of levels, below we carried
out test calculations to estimate the influence of the
substitution of $\sigma_{ph}$(NORAD) for the hydrogen-like
cross sections on the non-LTE results.

The model He I atom includes 183 allowed transitions.
The oscillator strengths $f_{ij}$ were taken
from NIST and the atomic structure calculations by
R. Kurucz\footnote{http://kurucz.harvard.edu/}.
For 14 transitions the radiative rates
are calculated with the Stark absorption profile from \citep[][He~I 4471~\AA]{he4471} and \citep[][the remaining 13 transitions]{Griem74}. We use the Voigt
absorption profile for the transitions with $f_{ij} >$ 0.01 in the range 900--6000~\AA\ and the Doppler one for the
remaining transitions.

{\it Transitions under the action of collisions with electrons.} 
For 196 bound-bound (b-b) transitions
between levels with $n \le 5$ we use the effective collision
strengths
$\Upsilon$ calculated by \citet{2000A&AS..146..481B} and \citet{1993ADNDT..55...81S}.
For the remaining
transitions we use the semiempirical formula from \citet{Reg1962} if the transition is allowed and take $\Upsilon$ = 1 if the transition is forbidden.

The electron-impact ionization cross sections are
calculated using the formula from \citet{seaton62}.

\subsection{Non-LTE Effects for He~I and Testing the Model Atom}\label{sect:test_nlte}

Before using the constructed model atom to analyze
the He I lines in HD 188101, we tested it by
determining the He abundance in chemically normal
stars for which non-LTE calculations were carried
out. We chose two stars with the solar chemical
composition from our previous papers: 21 Peg with \Teff\ = 10400~K, \lgg\ = 3.5 (10400/3.5), and a microturbulent velocity $v_{mic}$ = 0.5~\kms and $\iota$~Her with 17500/3.8 and $v_{mic}$ = 1.05~\kms. The
details of determining the atmospheric parameters
and the ESPaDOnS spectra\footnote{https://www.cadc-ccda.hia-iha.nrc-cnrc.gc.ca/en/search/} being used can be
found, for example, in \citep{2020MNRAS.499.3706M}.

\begin{figure*}[h!]
	\centering
	\includegraphics[scale=0.55]{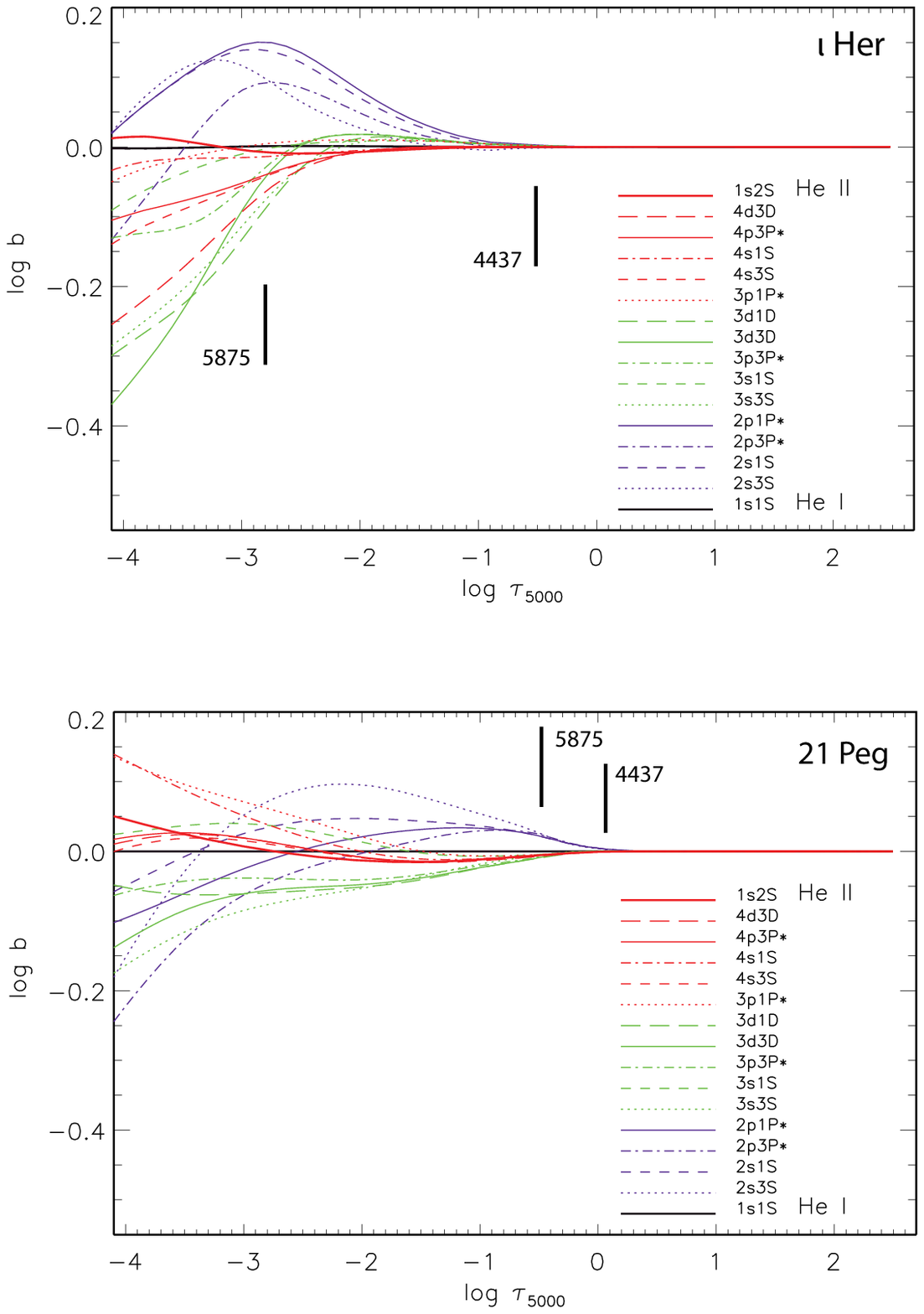}
	\caption{The $b$--factors of He I and He II levels as a function of the depth in the 17 500/3.8 ($\iota$~Her, the upper panel) and 10 400/3.5
		(21 Peg, the lower panel) model atmospheres. The formation depths of the He I 4437 and 5875 \AA\ line cores are indicated.}\label{pic:bfiHer21Peg}
\end{figure*}

Figure~\ref{pic:bfiHer21Peg} shows the $b$-factors of He I levels calculated
for the model atmospheres of 21 Peg and $\iota$~Her. In both models the neutral helium atoms dominate in
the total helium number density and, therefore, for the
ground He I state there are no departures from LTE ($b$ = 1) at all depths. The lower levels of those transitions
in which the observed lines are formed (Table~\ref{tab:testabunHe}), namely \eu{2s}{1}{S}{}{}, \eu{2p}{1}{P}{\circ}{} and \eu{2p}{3}{P}{\circ}{}, are overpopulated ($b_{low} >$ 1) relative to the equilibrium populations in
the line formation region, while the upper levels are
overpopulated, but to a lesser degree than the lower
ones ($b_{up} >$ 1, $b_{up} < $ $b_{low}$) for $\iota$~Her and underpopulated ($b_{up} <$ 1) for 21~Peg. The slight excess of the
photoionization rate compared to the photorecombination
rate for the ground state, the strong coupling
of highly excited states with the continuum, and the
spontaneous transitions to lower-lying levels are responsible
for this He I level population picture.

In the line formation region $b_{low} > 1$ and $b_{up}/b_{low} < 1$ for all of the transitions in Table \ref{tab:testabunHe}. The lines strengthen compared to LTE, and a lower
abundance is required to describe the observations.
Thus, the non-LTE abundance corrections $\Delta_{\rm NLTE} = \eps{NLTE} - \eps{LTE}$ are negative.

The abundances derived from individual lines in
21 Peg and $\iota$~Her are given in Table \ref{tab:testabunHe} for various
line formation scenarios -- LTE and non-LTE using
the Voigt (NLTE Voigt) and Stark (NLTE Stark)
profiles in our calculations of the radiative rates.
We also carried out non-LTE calculations using $\sigma_{ph}$(NORAD). The abundance difference between the
cases of $\sigma_{ph}$(OP) and $\sigma_{ph}$(NORAD) did not exceed
0.01 dex for any of the lines. This is because the
photoionization and photorecombination processes
associated with the He I \eu{4p}{3}{P}{\circ}{} level and higher levels
play a minor role in the mechanism of departures
from LTE. It turned out that the He I 4921, 5015, 5875, 6678, 7065 and 7282~\AA\ line profiles for $\iota$~Her could not be reproduced in the LTE calculations.
NLTE Voigt and NLTE Stark yield very close results
with a maximum abundance difference of 0.01 dex
for 21 Peg and 0.02 dex for $\iota$~Her. The non-LTE
corrections (NLTE Stark -- LTE) for individual lines
lie within the ranges from $-0.01$ to $-0.13$ dex for
21 Peg and from 0 to $-0.08$ dex for $\iota$~Her. For 21 Peg,
for which the LTE and non-LTE abundances were
determined for each of the lines, using the non-LTE
approach allows one to reduce the abundance error $\sigma = \sqrt{\sum(x-\bar{x})^2/(N_l - 1)}$ by a factor of 1.5. Here,
$N_l$ is the number of lines.

In the universally accepted scale we obtained
the non-LTE abundance $\eps{He}$ = 10.94$\pm$0.04 for 21~Peg and $\eps{He}$ = 10.95$\pm$0.03 for $\iota$~Her. These
values agree, within the error limits, with the standard
cosmic abundance $\eps{He}$ = 10.99$\pm$0.02 \citep{lodders21}.

\begin{table*}[th]
	\centering
	\caption {LTE and non-LTE helium abundances lg $N_{\rm He}/N_{\rm tot}$ in the atmospheres of the standard stars 21 Peg and $\iota$~Her. } \label{tab:testabunHe}
	\begin{tabular}{crlccccccc}
		\noalign{\smallskip} \hline
		&    &    &  \multicolumn{3}{c}{21 Peg} &  & \multicolumn{3}{c}{$\iota$ Her}    \\ 
		\cline{4-6}
		\cline{8-10}\noalign{\smallskip}
		$\lambda$(\AA) & lg~{\it gf} & \multicolumn{1}{c}{Transition} &  LTE     & NLTE   & NLTE   &  &  LTE   & NLTE   &  NLTE \\
		&         &               &          & Voigt  & Stark  &  &        & Voigt  & Stark \\
		\hline                                            
		4026   &  --0.37  & 2p$^3$P$^\circ$  -- 5d$^3$D  &  --1.08   & --1.10  & --1.10  &  & --1.10  &  --1.10 &  --1.10  \\
		4120   &  --1.47  & 2p$^3$P$^\circ$  -- 5s$^3$S  &  --1.06   & --1.08  & --1.08  &  & --1.05  &  --1.08 &  --1.09  \\
		4387   &  --0.89  & 2p$^1$P$^\circ$  -- 5d$^1$D  &  --1.11   & --1.12  & --1.12  &  & --0.99  &  --1.05 &  --1.05  \\
		4437   &  --2.02  & 2p$^1$P$^\circ$  -- 5s$^1$S  &  --1.06   & --1.07  & --1.07  &  & --1.04  &  --1.07 &  --1.08  \\
		4471   &    0.44  & 2p$^3$P$^\circ$  -- 4d$^3$D  &  --1.09   & --1.14  & --1.14  &  & --1.03  &  --1.09 &  --1.11  \\
		4713   &  --1.02  & 2p$^3$P$^\circ$  -- 4s$^3$S  &  --1.13   & --1.15  & --1.16  &  & --1.07  &  --1.14 &  --1.15  \\
		4921   &  --0.44  & 2p$^1$P$^\circ$  -- 4d$^1$D  &  --1.04   & --1.07  & --1.07  &  &  $::^1$ &  --1.02 &  --1.03  \\
		5015   &  --0.82  & 2s$^1$S   -- 3p$^1$P$^\circ$ &           &         &         &  &   $::$  &  --1.07 &  --1.09  \\
		5875   &    0.74  & 2p$^3$P$^\circ$  -- 3d$^3$D  &  --0.96   & --1.09  & --1.09  &  &   $::$  &  --1.07 &  --1.09  \\
		6678   &    0.33  & 2p$^1$P$^\circ$  -- 3d$^1$D  &  --0.99   & --1.11  & --1.11  &  &   $::$  &  --1.11 &  --1.12  \\
		7065   &  --0.20  & 2p$^3$P$^\circ$  -- 3s$^3$S  &  --0.93   & --1.03  & --1.03  &  &   $::$  &  --1.10 &  --1.12  \\
		7281   &  --0.84  & 2p$^1$P$^\circ$  -- 3s$^1$S  &           &         &         &  &   $::$  &  --1.06 &  --1.08  \\
		\hline
		&          & \multicolumn{1}{c}{Mean}        &  --1.05   & --1.10  & --1.10  &  & --1.05  &  --1.08 &  --1.09  \\
		&          & \multicolumn{1}{c}{$\sigma$(dex)}       &  ~~0.06   & ~~0.03  & ~~0.04  &  & ~~0.03  &  ~~0.03 &  ~~0.03  \\
		\hline \noalign{\smallskip}
	\end{tabular}
	
	$^1$ It is impossible to reproduce the observed profile.
	
\end{table*}


{\it Comparison with other authors.} The departures
from LTE in He I lines were investigated in
many papers \citep[see, e.g.,][]{1973ApJS...25..433A, 1980A&A....81....8D, 1989A&A...222..150H, 1990SvAL...16..231S, 1994PASJ...46..181T, 1995A&A...293..457L, 2005A&A...443..293P, 2018AstL...44..621K}. A direct comparison with the results of \citet{2018AstL...44..621K}, who performed
their calculations for 21 Peg and $\iota$~Her using the
same atmospheric parameters as those in our paper,
can be made. For 21 Peg and the He I 5875 and 6678~\AA\ lines with maximum departures
from LTE \citet{2018AstL...44..621K} obtained
amore negative non-LTE correction, $\Delta_{\rm NLTE} = -0.17$~dex compared to our $\Delta_{\rm NLTE} = -0.13$ and $-0.12$~dex. As a consequence, for both stars \citet{2018AstL...44..621K} obtained a lower mean He
abundance, by 0.04~dex.

\section{DETERMINATION OF $T_{\rm eff}$ AND \lgg}\label{sect:atm}

We determined $T_{\rm eff}$ from photometry and \lgg\ from
Balmer hydrogen lines and evolutionary tracks. The
interstellar extinction data were extracted from various
maps \citep[][-- Arenou+1992, Green+2019, Gontcharov 2017, Lallement+2022, respectively]{1992A&A...258..104A, 2019ApJ...887...93G, 2017AstL...43..472G, 2022A&A...661A.147L}. From the Green+2019
maps\footnote{http://argonaut.skymaps.info/}  we determined the color excess $\rm E(g-r) = 0.050^{+0.054}_{-0.050}$, using the Bayestar2019 version. For the
conversion to the color excess in the Johnson system
we took the coefficients suitable for the object under
study from \citet{1996AJ....111.1748F}, E(B-V) = E(g-r)/1.05. The Lallement+2022 maps\footnote{https://explore-platform.eu} give $\rm A_V = 0.067^{+0.032}_{-0.019}$. Using the interstellar extinction
law $\rm R_V \times A_{\lambda}/A_V$ with $\rm R_V = 3.1$ from \citet{1990ARA&A..28...37M}, we found the color excess E(B-V). It can be seen
from Table~\ref{tab:tabebmv} that different maps give color excesses
from 0.02 to 0.08. To calculate the uncertainty in
the color excess, we took into account the error in E(B-V) given by the maps and the uncertainty in
the distance to the star (see Table \ref{tab:param}).

\begin{table*}[ht]
	\centering
	\caption {The color excesses E(B-V) determined from
		four interstellar extinction maps} \label{tab:tabebmv}
	\medskip
	\begin{tabular}{c|c|c|c}
		\hline
		Arenou+1992         &  Green+2019            &  Goncharov 2017         & Lallement+2022          \\ \hline
		$0.051^{+0.065}_{-0.049}$  &  $0.048^{+0.051}_{-0.048}$ &   $0.081^{+0.041}_{-0.016}$ & $0.022^{+0.010}_{-0.006}$   \\ \hline
	\end{tabular}\\
\end{table*}

\begin{figure*}[h!]
	\centering
	\includegraphics[scale=0.2]{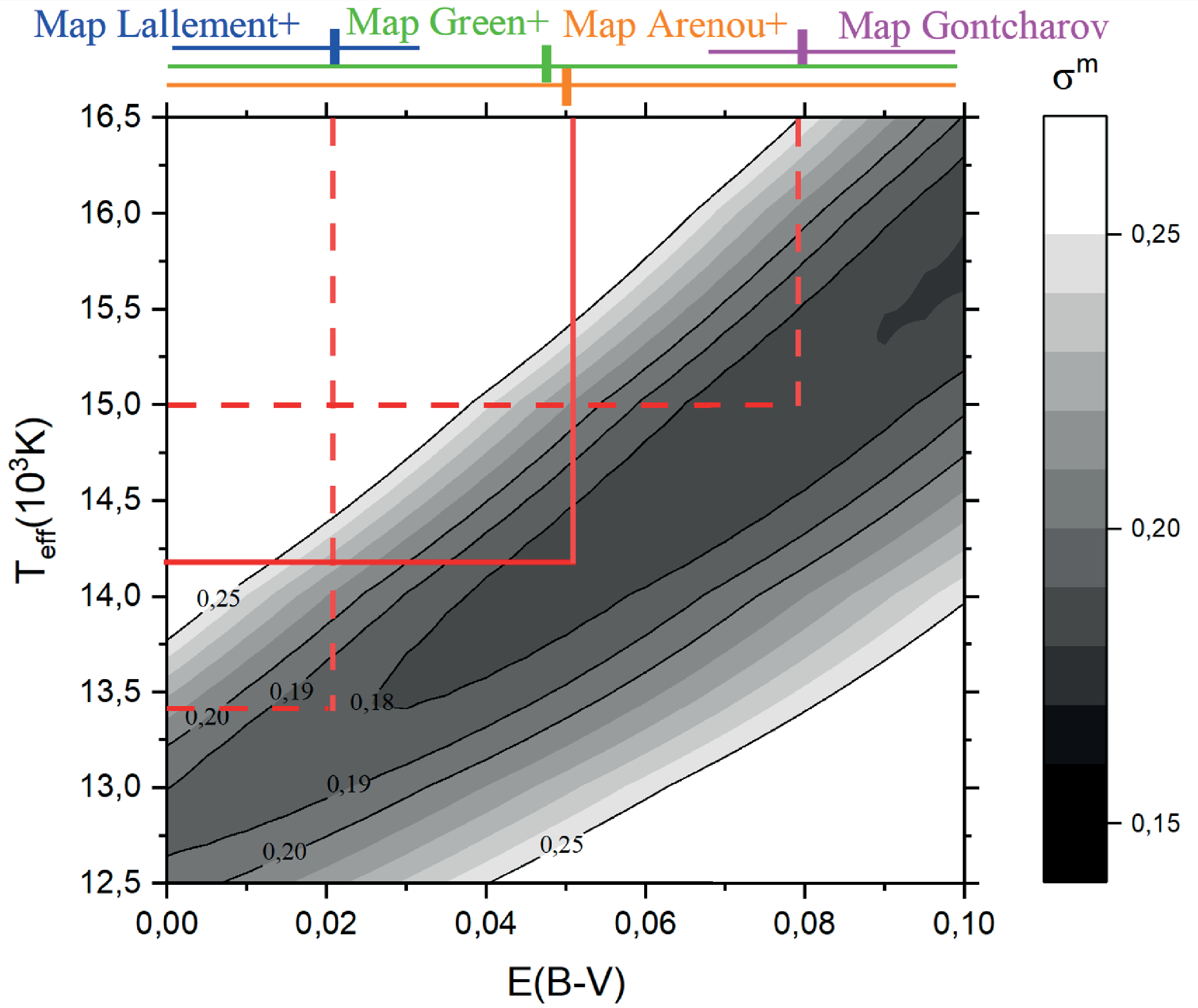}
	\caption{The deviations of the theoretical fluxes from the observed ones (in magnitudes) for different $T_{\rm eff}$ and E(B-V) are
		indicated by the shading with different intensities. The solid and dashed red straight lines mark the derived E(B-V) and $T_{\rm eff}$
		with their errors.} \label{pic:teffbmv}
	\includegraphics[scale=0.35]{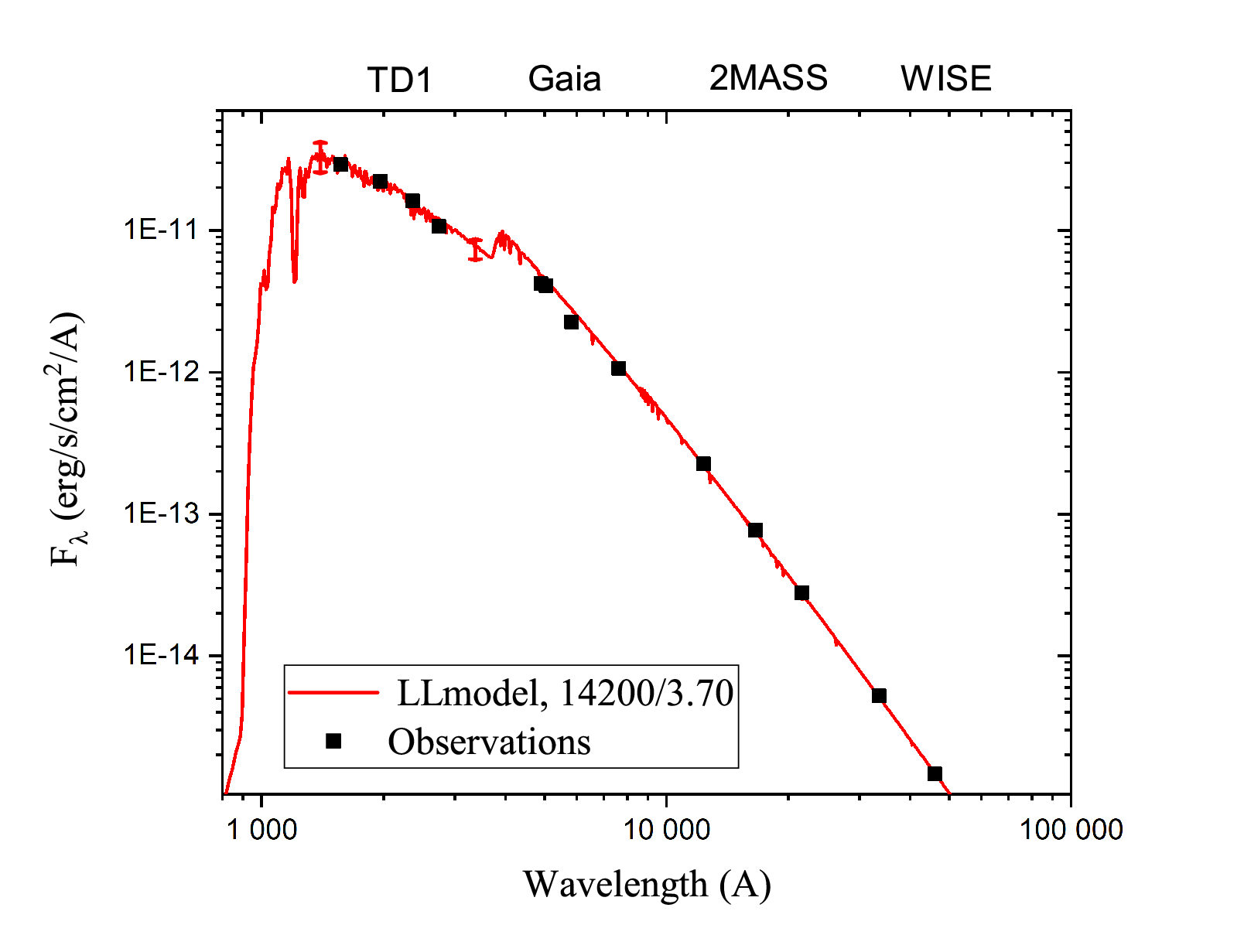}
	\caption{The observed fluxes corrected for interstellar extinction (squares) and the theoretical fluxes smoothed by a Gaussian
		with a FWHM of 10~\AA\ (curve) for the model with the ultimately adopted atmospheric parameters. The bar of the theoretical
		flux indicates the error due to the uncertainty in the effective temperature $\sigma_O$.}
	\label{pic:sed188}
\end{figure*}

$T_{\rm eff}$ and the angular diameter $\theta$ were determined
by minimizing the deviations of the theoretical
fluxes $F_{\lambda}$ convolved with the filter transmission
curves\footnote{https://svo2.cab.inta-csic.es/theory/fps3/index.php?mode=voservice}, from the extinction-corrected observed
fluxes. The fluxes $F_{\lambda}$ were computed by the LLmodels
code for a set of $T_{\rm eff}$ at fixed \lgg\ = 4.0 and solar
chemical composition. Figure~\ref{pic:teffbmv} shows the deviations
from the observations in magnitudes, $\sigma^m =\sqrt{\sum_{i=1}^{N_{obs}}(m_i^{syn} - m_i^{obs})^2/N_{obs}}$ for a set of E(B-V) and $T_{\rm eff}$, and the color excesses with their errors
corresponding to HD 188101 on different maps are
marked. The minimum deviation $\sigma^m = 0.17$was found in the ranges of E(B-V) from 0.09 to 0.16 and $T_{\rm eff}$ from 15500 to 18000 K. However, in view of the degeneracy in E(B-V) and $T_{\rm eff}$ we fixed E(B-V)=0.05$\pm$0.03 based on the interstellar extinction
maps. We determined $T_{\rm eff}$, $\theta$ and the errors related to
the uncertainties in the color excess $\sigma_E$ and surface
gravity $\sigma_g$. The errors attributable to the observations $\sigma_O$ were estimated by the Monte Carlo method for $T_{\rm eff}$ and by calculating the standard deviation from the
observations for the angular diameter. The total error
is calculated as $\sigma_{atm}^2 = \sigma_E^2+\sigma_g^2+\sigma_O^2$.
In addition,
we estimated the influence of the overabundances
of some chemical elements ([Si/H] $\simeq$ 1.0, [Ti/H] $\simeq$ 1.0, [Fe/H] $\simeq$ 0.5), found by us in HD 188101 
(see
 Section~\ref{subsecchabun}) on the determination of \Teff\ and $\theta$. The
 fluxes $F_{\lambda}$ were calculated in the model with a modified chemical composition, which led to a change in \Teff\ by +40~K and $\theta$ by -0.7~$\mu as$ ($\sigma_{abun}$ in Table \ref{tab:teffthetasig}). The final values of \Teff\ and the angular diameter
 corresponding to E(B-V) = 0.05, and the errors
 in the stellar parameters are given in Table~\ref{tab:teffthetasig}.
In Fig.~\ref{pic:sed188} the theoretical fluxes for the ultimately adopted
model are compared with the observations. Note that
with the derived color excess and \Teff\ the deviation $\sigma^m = 0.18$ does not differ greatly from the minimum $\sigma^m = 0.17$.

\begin{table*}[th]
    \centering
	\caption {The errors in \Teff, the angular diameter, and \lgg\
		for HD188101 and the final values of the stellar parameters} \label{tab:teffthetasig}
	\medskip
	\begin{tabular}{lrr|lr}
	\hline
	& $T_{\rm eff}$~(K) &   $\theta\ (\mu as)$  & & \lgg\  \\  \hline
	           & 14200   &   67.5      & & 3.70          \\  
	$\sigma_E$    & 800     &   1.7    & $\sigma_{o}  $ & 0.05       \\  
	$\sigma_g$    & 250     &   1.0    & $\sigma_{T}  $ & 0.15        \\  
	$\sigma_O$    & 520     &   0.5    & $\rm \sigma_{LTE}$ & --0.10   \\  
    $\sigma_{abun}$    & +40     &   --0.7    & $\sigma_{abun}$ & +0.02   \\  
	$\sigma_{atm}$ & 990     &   2.0    & $\sigma_{atm}$ & 0.16       \\  \hline
	\end{tabular}
\end{table*}

\begin{figure*}[h]
	\hspace{-15mm}	
	\includegraphics[scale=0.45]{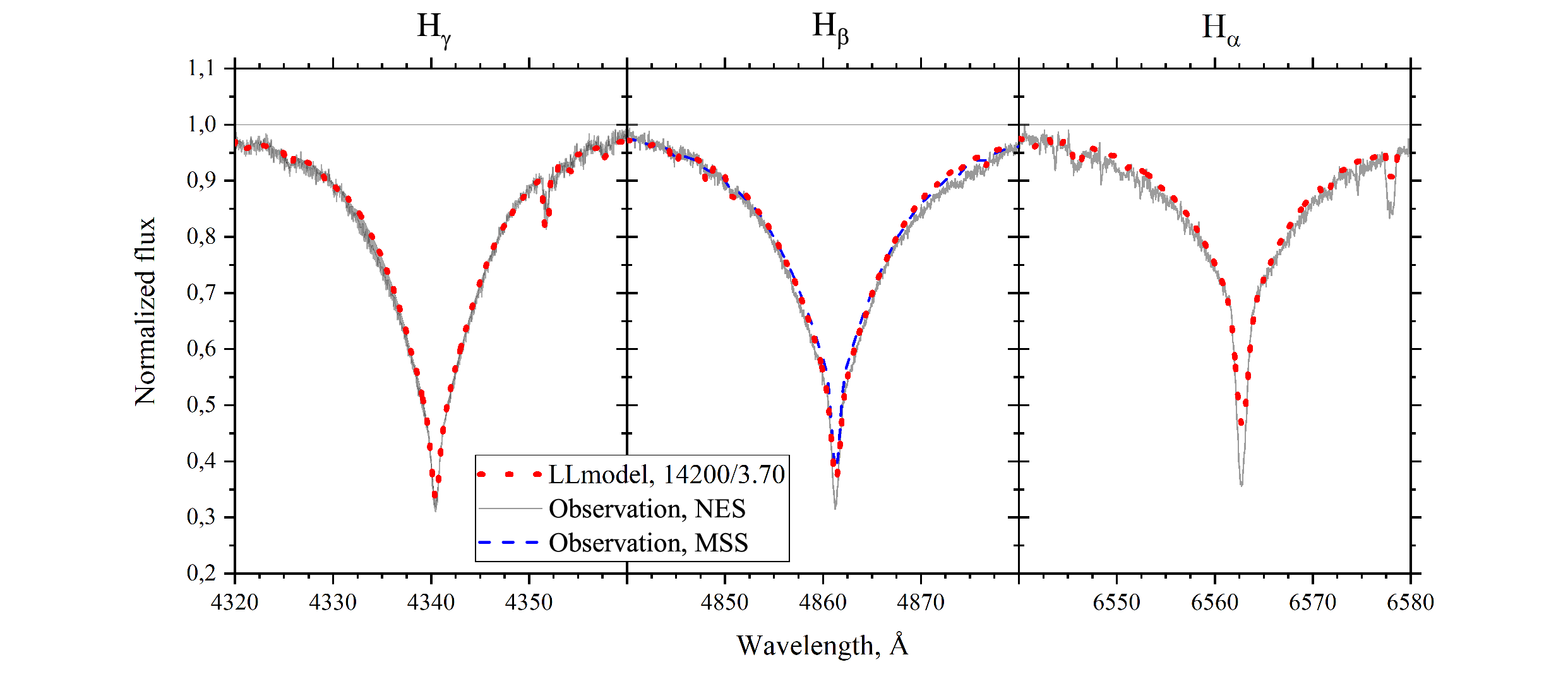}
	\caption{The theoretical hydrogen line profiles for the model with  \Teff\ = 14200~K and \lgg\ = 3.70 in comparison with the observed NES (solid curve) and MSS (dashed curve) spectra for HD 188101.}
	\label{pic:Balm188}
\end{figure*}

Let us now determine \lgg\  from the  H~I lines
whose wings are sensitive to \lgg\ at $T_{\rm eff} > 10000$~K.
We computed a grid of model atmospheres using
the LLmodels code and He I line profiles using the
SynthVb code for a set of $T_{\rm eff}$ and \lgg. The $\rm H_{\alpha},\ H_{\rm \beta},\ H_{\gamma}$ line profiles are in best agreement with the
observations at \lgg\ = 3.70 (Fig.~\ref{pic:Balm188}).

The uncertainties in the derived \lgg\ are attributable
to the errors in the observations $\sigma_{o}$ and the
effective temperature $\sigma_{T}$ (Table~\ref{tab:teffthetasig}). We also estimated
the systematic error in \lgg due to the neglect of
the departures from LTE, $\rm \sigma_{LTE}$. For this purpose,
we compared the $\rm H_{\beta}$ profiles computed when abandoning
the assumption of LTE using the TLUSTY
code  \citep{2007ApJS..169...83L} with \Teff\ = 15000~K and \lgg\ = 4.0 with our LTE calculations for the
same atmospheric parameters. Compared to non-
LTE, the LTE profile is stronger in the wings and,
therefore, the same observed profile is reproduced in
our LTE calculations at a lower value of \lgg, by 0.1~dex. In addition, using the LLmodels models with
abundances typical for HD 188101 ([Si/H] $\simeq$ 1.0, [Ti/H] $\simeq$ 1.0, [Fe/H] $\simeq$ 0.5) leads to a slight increase
in  \lgg\ by 0.02 dex ($\sigma_{abun}$ in Table~\ref{tab:teffthetasig}).

We also determined \lgg\ for HD~188101 using
MESA (Modules for Experiments in Stellar Astrophysics)
evolutionary tracks \citep[Modules for Experiments in Stellar Astrophysics,][]{2016ApJS..222....8D, 2016ApJ...823..102C}.
Assuming the initial chemical composition
to be solar, neglecting the rotation and
the magnetic field, and using our estimates of the
luminosity $L/L_{\odot}  = 360^{+640}_{-200}$ and $T_{\rm eff}$, we estimated \lgg\ = $4.03^{+0.31}_{-0.35}$ and the mass $M/M_{\odot} = 4.13\pm 0.77$. The errors were found by the Monte Carlo
method with specified normal lg~$L/L_{\odot},\ T_{\rm eff}$ distributions.
Both methods lead to consistent values
of \lgg, within the error limits.

We did not find a sufficient number of weak and
strong lines in the observed spectrum to determine
the microturbulent velocity $v_{mic}$ and took a value typical
for CP stars, $v_{mic}$ = 2.0~\kms. 

The parameters of HD 188101 derived in this paper
and \vsini\ and \Per\ taken from \citet{2023AstBu..78..141Y} are presented in Table~\ref{tab:param}. The inclination of the rotation
axis to the line of sight was calculated using \vsini, \Per\ and the radius.

\begin{table}
	\centering
	\caption {Parameters of the star HD~188101.} \label{tab:param}
	\medskip
	\begin{tabular}{lr}
		\hline
		$T_{\rm eff}$(K) &  $14200\pm 990$       \\
		$\theta\ (\mu as)$ & $67.5\pm 2.9$ \\
		$R/R_{\odot}$ & $3.1^{+2.1}_{-0.9}$ \\
		\lgg\ (CGS) & $3.70\pm0.16$ \\
		$M/M_{\odot}$ & $4.13\pm 0.77$ \\
		$v_{mic}$ (\kms) & 2.0 \\
		E(B-V) & $0.052\pm 0.30$ \\
		r (pc) & $430^{+280}_{-120}$ \\
		$B_p$ (kG) & $<$3.0 \\
		\vsini (\kms) & 33 \\
		$i\ (^{\circ})$ & $42^{+19}_{-15}$ \\
		\Per\ (days) & 3.98726 \\
		 \hline
	\end{tabular}
\end{table}

\section{CHEMICAL COMPOSITION} \label{subsecchabun}

\begin{figure*}[h]
	\hspace{-15mm}
	\includegraphics[scale=1.0]{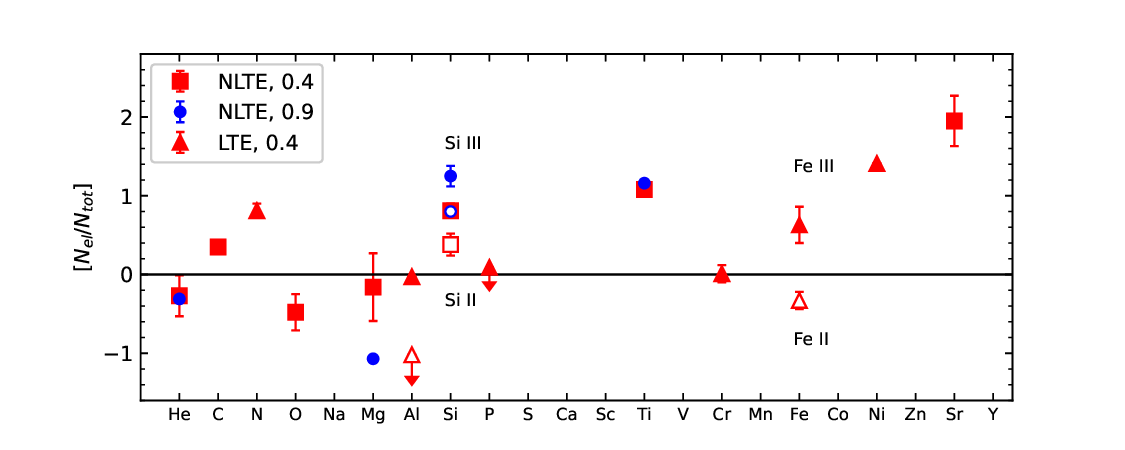}
	\caption{The chemical composition of HD 188101. The Al, Si, and Fe abundances from the lines of the II (unfilled symbols) and
		III (filled symbols) ionization stages are given. The arrows indicate the upper limits. The error bar indicates the dispersion of
		the abundance when averaged over all lines of the element.}\label{pic:Abunhd188}
\end{figure*}

We determined the abundances from individual
lines based on the MSS spectra and the NES spectrum
for the lines uncovered by the MSS spectra.
For the He~I, Mg~II, Si~II, Si~III and Ti~II 
lines we made the determinations for two rotation phases,
0.4 and 0.9, based on the average spectra. In the
first iteration we used the 14200/3.70 with $v_{mic}$=2 \kms and the solar chemical composition \citep{lodders21}, but it was subsequently recalculated
using the abundances of individual elements (the Si,
Ti, and Fe overabundances) derived in the first iteration.
Our results are presented in Table~\ref{tab:abun188}. Figure~\ref{pic:Abunhd188} shows the mean LTE and non-
LTE abundances \Neltot for phases 0.4 and 0.9 (if
the data are available). The He~I 4471~\AA\ line was not
used in calculating the mean. To calculate \Neltot, we used the protosolar abundances of elements from \citet{lodders21}. 

\begin{longtable}{lcc|lll|lll}
	\caption {LTE (L) and non-LTE (N) abundances $\rm lg(N_{el}/N_{tot})$ from individual lines in HD 188101 (14200/3.70 model)
		at two phases (0.4, 0.9) from the MSS spectra. \Eexc\ is the lower-level excitation energy \label{tab:abun188} }\\
	\hline
		Line    & \Eexc, & lg gf & \multicolumn{3}{c}{Phase 0.4}  &  \multicolumn{3}{c}{Phase 0.9}  \\
		& eV & & ~~~L  &  ~~~N  & \Neltot &  ~~~L   &  ~~~N & \Neltot \\ 
		\hline
		He~I 4437    & 21.21 &  -2.03 & -1.07     &-1.09     &-0.04      &          &         &    \\ 
		He~I 4713    & 20.96 &  -1.03 & -1.20     &-1.27     &-0.22      &   -1.30  &   -1.36 & -0.31   \\
		He~I 5015$^1$& 20.62 &  -0.82 & -1.51     &-1.61     &-0.56      &          &         &    \\
		C~II 4267*   & 18.05 &   0.96 & -3.10     &-3.13     &0.35      &          &         &    \\
		N~II 4607    & 18.46 &  -0.52 & -3.27     &          &0.83      &          &         &    \\
		N~II 4613    & 18.47 &  -0.69 & -3.32     &          &0.78      &          &         &    \\
		N~II 4630    & 18.48 &   0.08 & -3.38     &          &0.72      &          &         &    \\
		N~II 4643    & 18.48 &  -0.37 & -3.17     &          &0.93      &          &         &    \\
		Mean      &       &        & -3.29(09) &          &0.81  &          &         &    \\
		O~I 5329*    & 10.74 &  -1.24 & -3.32     &-3.43     &-0.21      &          &         &    \\
		O~I 6155*    & 10.74 &  -0.66 & -3.67     &-3.84     &-0.62      &          &         &    \\
		O~I 6158*    & 10.74 &  -0.30 & -3.65     &-3.83     &-0.61      &          &         &    \\
		Mean      &       &        & -3.55(20) &-3.70(23) &-0.48 &          &         &    \\
		Mg~II~4427   & 10.00 &  -1.21 & -4.37     &-4.34     &0.09       &          &         &    \\
		Mg~II~4433   & 10.00 &  -0.91 & -4.39     &-4.35     &0.08       &          &         &    \\
		Mean      &       &        & -4.38(01) &-4.35(01) &0.08  &          &         &    \\
		Mg~II~4481   &  8.86 &   0.99 & -4.97     &-5.09     &-0.66 & -5.48    &-5.50    &-1.07\\
		Al~II 4663   & 10.60 &  -0.28 & <-6.59    &          &<-1.02      &          &         &    \\
		Al~II 6231*  & 13.07 &   0.52 & <-6.59    &          &<-1.02      &          &         &    \\
		Al~II 6243*  & 13.08 &   0.74 & <-6.59    &          &<-1.02      &          &         &    \\
		Al~III 4512  & 17.81 &   0.41 &  -5.60    &          &-0.03      &          &         &    \\
		Si~II 4673   & 12.84 &  -0.34 & -4.36     &-4.10     &0.34  &-3.94     &-3.64    &0.80\\
		Si~II 5041*  & 10.07 &  0.15  & -4.78     &-3.87     &0.57  &          &         &    \\
		Si~II 5055*  & 10.07 &  0.53  & -5.06     &-4.21     &0.23  &          &         &    \\
		Mean      &       &        & -4.73(29) &-4.06(14) &0.38  &          &         &    \\
		Si~III 4552  & 19.02 &   0.18 & -3.44     &-3.64     &0.80  &-2.97     &-3.25    &1.19\\
		Si~III 4567  & 19.02 &  -0.04 & -3.56     &-3.71     &0.73  &-3.05     &-3.28    &1.16\\
		Si~III 4574  & 19.02 &  -0.51 & -3.44     &-3.54     &0.90  &-2.87     &-3.04    &1.40\\
		Mean      &       &        & -3.48(07) &-3.63(09) &0.81  & -2.96(09)&-3.19(13)&1.25\\
		P~II 4499    & 13.38 &   0.61 & <-6.43    &          &<0.09 &          &         &    \\
		Ti~II 4443   & 1.08  &  -0.71 &-6.18      &-5.98     &1.07      &          &         &    \\
		Ti~II 4501   & 1.12  &  -0.77 &-6.05      &-5.86     &1.19      &          &         &    \\
		Ti~II 4563   & 1.22  &  -0.69 &-6.27      &-6.05     &1.00      &  -6.11   & -5.89   &1.16    \\
		Ti~II 4571   & 1.57  &  -0.31 &-6.17      &-6.00     &1.05      &  -6.05   & -5.89   &1.16    \\
		Mean      &       &        & -6.17(08) &-5.97(07) &1.08      &          &         &    \\
		Cr~II 4558   & 4.07  &  -0.37 & -6.34     &          &-0.02       &          &         &    \\
		Cr~II 4588   & 4.07  &  -0.65 & -6.44     &          &-0.12       &          &         &    \\
		Cr~II 4634   & 4.07  &  -1.05 & -6.29     &          &0.03        &          &         &    \\
		Cr~II 4824   & 3.87  &  -0.92 & -6.17     &          &0.15        &          &         &    \\
		Mean      &       &        & -6.31(11) &          &0.01        &          &         &    \\
		Fe~II 4508   & 2.85  &  -2.09 & -4.89     &          & -0.39      &          &         &    \\
		Fe~II 4522   & 2.84  &  -2.03 & -4.94     &          & -0.44      &          &         &    \\
		Fe~II 4555   & 2.83  &  -2.16 & -4.84     &          & -0.34      &          &         &    \\
		Fe~II 4583   & 2.81  &  -1.86 & -4.72     &          & -0.22      &          &         &    \\
		Mean      &       &        & -4.83(11) &          & -0.33      &          &         &    \\
		Fe~III 4431  & 8.25  &  -2.57 & -3.83     &          & 0.67       &          &         &    \\
		Fe~III 5127* & 8.66  &  -2.03 & -3.66     &          & 0.84       &          &         &    \\
		Fe~III 5156* & 8.64  &  -1.99 & -4.11     &          & 0.39       &          &         &    \\
		Mean      &       &        & -3.87(23) &          & 0.63       &          &         &    \\
		Ni~II 5050*  & 12.46 &   0.24 & -4.34     &          & 1.41       &          &         &    \\
		Sr~II 4077*  & 0.00  &   0.14 & -7.43     &-7.34     & 1.73 &          &         &    \\
		Sr~II 4215*  & 0.00  &  -0.17 & -6.98     &-6.89     & 2.18 &          &         &    \\
		Mean      &       &        & -7.21(32) &-7.12(32) & 1.95&          &         &    \\
		\hline
	\multicolumn{9}{l}{In parenthesis 100$\cdot \sigma$(dex).} \\
	\multicolumn{9}{l}{$^1$ From the MSS spectra at phases 0.1-0.3.} \\
	\multicolumn{9}{l}{*From the NES spectrum.} \\
\end{longtable}

In addition to the random errors $\sigma$ due to the
scatter in the abundances from different lines given
in Table~\ref{tab:abun188}, we estimated the changes in the mean
abundance due to the error in \Teff. Our LTE (or non-
LTE, if the model atom is available) calculations were
carried out for the model with \Teff\ = 15200~K without any change in other atmospheric parameters.
An increase in \Teff\ by 1000~K led to a change in the
abundance by $-0.20$, $-0.26$, and $-0.27$ dex for He~I 4437~\AA, 4713~\AA\ and 5015~\AA,$-0.33$~dex for N~II, by +0.12 dex for Mg~II 4427 and 4433~\AA, by +0.16~dex for Mg~II 4481~\AA, by +0.26~dex for Si~II, by $-0.34$~dex for Si~III, by +0.32~dex for Fe~II, and by $-0.09$~dex for Fe~III.

\begin{figure*}[th]
	\centering
 	\includegraphics[scale=0.45]{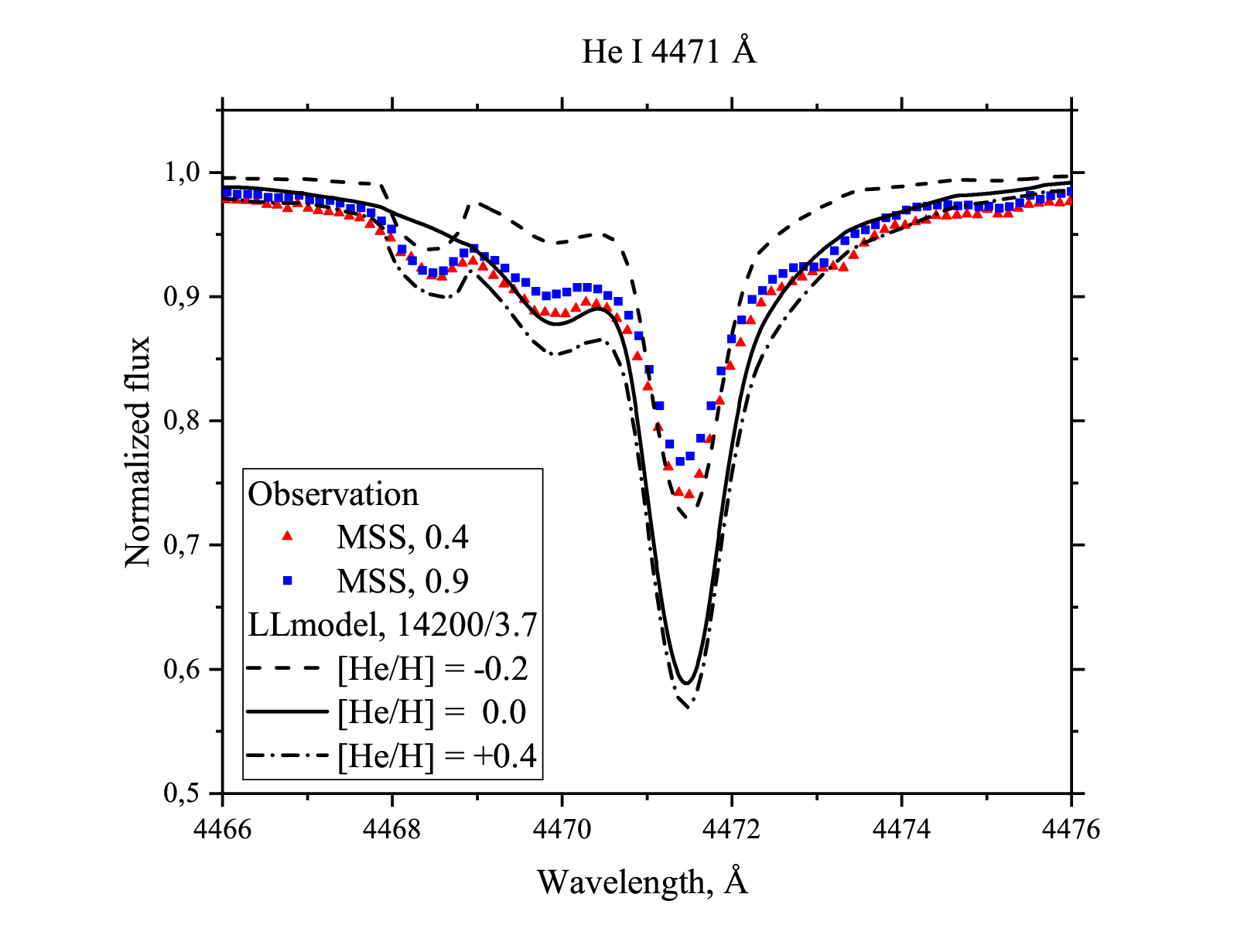}
 	\caption{Demonstration of the impossibility of describing the He~I 4471~\AA\ profile (the triangles and squares represent the MSS
 		spectra at phases 0.4 and 0.9, respectively) within the framework of a homogeneous model, without taking into account the
 		stratification and spots. The synthetic spectra were constructed in non-LTE with various [He/H] abundances.}\label{pic:He4471hd188}
 \end{figure*}

It turned out to be impossible to reproduce the
He~I 4471~\AA\ and 4921~\AA\ line profiles neither under the
assumption of LTE nor when abandoning it. Figure \ref{pic:He4471hd188}
shows the theoretical profiles that agree best with
the observations in the line wings and core for
He~I 4471~\AA . The He abundance difference between these
two cases is 0.6 dex. The profile computed with the
solar abundance is shown for comparison. We derived
the solar abundance from the weak Mg~II 4427 and 4433~\AA\ lines and an underabundance of $-0.74$ dex
from the strong Mg~II 4481~\AA\ line. For Si, Fe, and
Al in the LTE approximation the abundances from
the lines of the II and III ionization stages differ by
$\simeq 1$~dex. The Si II and Si III lines have non-LTE
corrections of opposite signs and, therefore, the abundance
difference between Si II and Si III decreases in
our non-LTE calculations. It is completely removed
by raising \Teff\ by $\simeq$700~K, which is within the temperature
error limits. We suggest that a non-LTE
analysis is needed to reconcile the abundances from
the Al II and Al III lines and from the Fe II and Fe III
lines.

Large Si, Ti, and Sr overabundances are observed
in HD 188101, and there is a hint at a He underabundance.
Note that no criterion has been established in
the literature for the He underabundance in He-weak
stars. Based on such observational manifestations
typical for He-weak stars as Si, Ti, and Sr overabundances,
a possible slight He underabundance, and a
weak magnetic field, we classify HD 188101 as a He-weak
SiTiSr star.

\section{DISCUSSION}\label{sect:discus}

\begin{figure}[th!]
	\centering
	\includegraphics[scale=0.40]{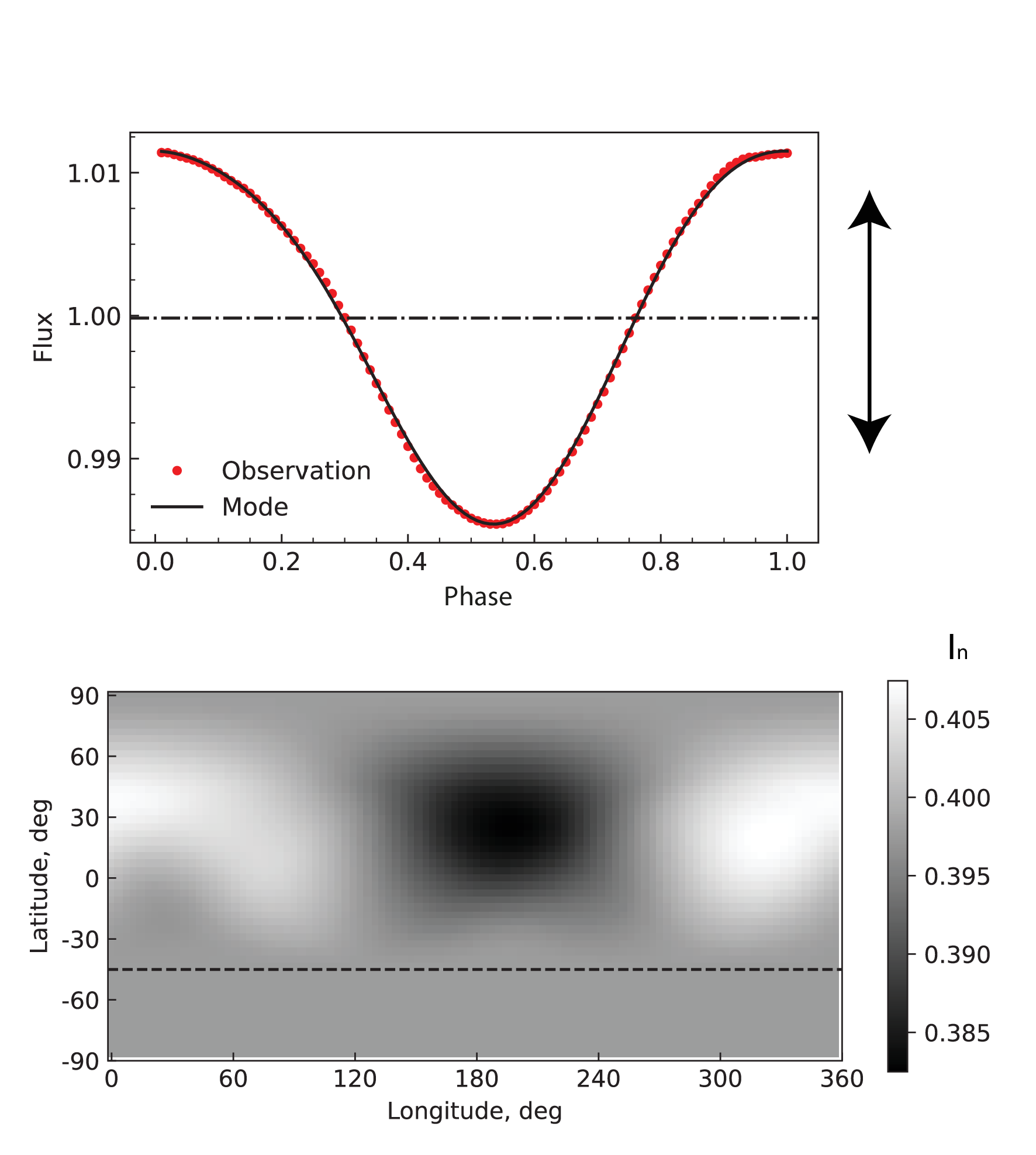}
	\caption{Upper panel: the light curve of HD 188101 in the Kepler filter (dots) in comparison with the model light curve (solid line). Lower panel: the derived intensity distribution.
	}
	\label{fig:varcont}
\end{figure}

The change in line absorption with rotation phase,
the different abundances from different lines of the
same species for the same phase, and the impossibility
of reproducing the He~I 4471 and 4921~\AA\ line profiles within the classical model atmosphere suggest that the surface layers are inhomogeneous in
chemical composition. We assumed the presence of
spots on the surface that create the observed features
due to the stellar rotation. Based on the light curve
(Fig.~\ref{fig:varcont}, the upper panel), we found the distribution
of the intensity emerging along the normal over the
stellar surface by the Tikhonov regularization method
(Fig.~\ref{fig:varcont}, the lower panel). The dashed line marks
the boundary below which the stellar surface is not
observed. For details, see, e.g., \citet[][]{2013ARep...57..548K}.
Spots with different intensities $\simeq 60^{\circ}$ in size typical for
CP stars can be seen.

With the help of the LLmodels code and the convolution
with the Kepler transmission curve, we computed
the corresponding fluxes for phases 0.4 and 0.9
using the previously derived elemental abundances
for them. The right arrow on the upper panel of
Fig.~\ref{fig:varcont} indicates the difference of these fluxes. It can
be seen that the scales of the brightness variations
correspond to the flux variations due to the abundance
variations. Note that the change in absorption beyond
the ionization threshold of the ground Si~I level ($\lambda$ = 1528~\AA) plays the main role in the flux variations. The
flux in the far ultraviolet decreases with increasing
Si abundance and increases in a longer-wavelength
range.

\begin{figure}[th]
	\centering
	\includegraphics[scale=0.72]{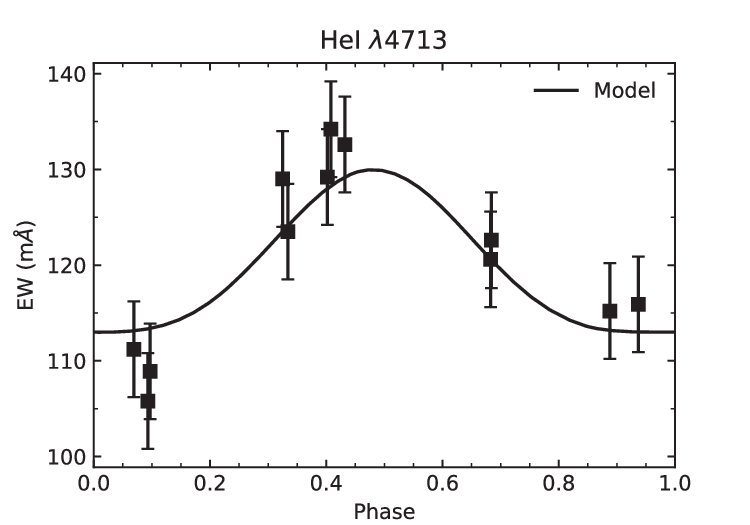}
	\caption{ Change in the He~I 4713~\AA\ equivalent widths with rotation phase (squares). The curve corresponds to the model with a circular equatorial spot.
	}
	\label{fig:modelHe4713}
\end{figure}

Consider the variability of the He~I 4713~\AA\ line
equivalent width within the simplest model. Assuming
the spot to be an equatorial spherical circle and
the fluxes in the continuum of the spot and outside
it to be identical, we estimated the spot sizes and
position on the surface (the solid curve in Fig. \ref{fig:modelHe4713})
and the [He/H] abundance in (from 0.0 to +0.1) and
outside (from $-0.2$ to $-0.4$) the spot. Just as from the
light curve, we obtained a spot size typical for peculiar
stars, $\simeq 60^{\circ}$. This analysis points to the need for applying
more complex models, such as those in \citet{2024Univ...10..341P, 2024MNRAS.52710376P}. The element on the stellar surface being resolved by MSS
has an extent in longitude $\Delta l^{\circ} = 90^{\circ}\times {\rm c/R}/V{\rm sin}i \simeq 60^{\circ}$. Therefore, observations using an instrument
with a resolution R $>$ 60 000 is required for Doppler
mapping.

Thus, a nonuniform distribution of chemical elements
over the surface, typical for peculiar stars,
is most likely responsible for the variability of the
brightness and absorption in spectral lines. Within a
simple model we obtained the He abundance spot and
the intensity at phase 0.4. Our abundance analysis at
this phase led to an elemental abundance closer to the
solar one compared to that at phase 0.9, where the Si
spot is located. 

The He~I 4471 and 4921~\AA\ line profiles are not
reproduced irrespective of the phase. The non-LTE
effects cannot be responsible for the strong wings
and the weak core of the observed profile relative to
the theoretical one. This may argue for vertical He
stratification or a more complex spot structure.

\section{CONCLUSIONS}

Based on our analysis of the spectroscopic time
series and photometric data, we determined the
atmospheric parameters of the star HD 188101
with a weak magnetic field \citep[$|B_{z}| <$ 1~kG,][]{2023AstBu..78..141Y}: \Teff\ = 14200 $\pm990$~K and \lgg\ = 3.70 $\pm0.16$. The surface gravity derived from the
profiles of Balmer H I lines agrees, within the error
limits, with \lgg\ derived from evolutionary tracks.
The profiles are well described at \Teff\ determined from
photometric observations, suggesting that E(B-V)
obtained from interstellar extinction maps is correct.
The parameters derived by us are consistent with the
results of \citet{2023AstBu..78..141Y}, 14700/3.8.

We constructed a model He I atom. Even our
non-LTE calculations do not allow the He I 4471
and 4921~\AA\ line profiles to be reproduced within the
framework of a classical, homogeneous model atmosphere,
while the He I 4437, 4713, and 5015~\AA\ lines
give different abundances for the same rotation phase,
but in all of the cases there is an underabundance
of He relative to its solar abundance, from $-0.04$ to
$-0.56$~dex.

We carried out non-LTE calculations for C II, O I,
Mg II, Si II?Si III, Ti II, and Sr II. The Mg II 4427
and 4433~\AA\ lines give a nearly solar abundance, while
Mg II 4481~\AA\ gives a value lower by 0.74 dex. The
non-LTE abundances from the Si II and Si III lines
can be reconciled if we raise \Teff\ by $\simeq$700~K, which is
within the temperature error limits. Overabundances
of Si, Ti, and Sr with [Si/H] $\simeq$ 1.0, [Ti/H] = 1.12 and [Sr/H] = 1.95 were revealed in the star.

Thus, judging by its chemical composition,
HD 188101 belongs to the Bp class, the group
of He-weak SiTiSr stars, and shows photometric
variability and absorption in the He I, Mg II, Si II,
Si III, Ti II, and Fe II lines typical for peculiar stars.
The changes of the absorption in the He I and Mg II
4481 \AA\ lines anticorrelate with the absorption in the
Si II, Si III, Ti II, and Fe II lines.

To understand the causes of the photometric and
spectroscopic variability of HD 188101, to reproduce
the profiles of strong He I lines, and to remove the
discrepancies in the abundances from different lines of
the same element, it is necessary to model the surface
layers and radiative transfer by taking into account
the chemical inhomogeneity and to obtain a series of
high-resolution spectroscopic observations uniformly
distributed in rotation phases.

{\bf Acknowledgments.} The work presented in Subsections \ref{sect:NLTE} and \ref{sect:test_nlte}
	was performed within RSF project no. 23-12-00134.
	The work performed in Section \ref{sect:obs} was supported by
	RSF grant no. 25-12-00003. We used the NIST,
	NORAD, Simbad, TOPbase, VALD, and ADS\footnote{http://adsabs.harvard.edu/abstract\_service.html}
	databases. We are grateful to I. S. Potravnov for the
	consultations in working with the spectroscopic time
	series.




\end{document}